 \documentstyle[emulateapj,apjfonts,psfig]{article}

\newcommand{\LCDM}{$\Lambda$CDM}

\newcommand{\hkpc}{$h^{-1}{\ }{\rm kpc}$}

\newcommand{\hMsun}{$h^{-1}{\ }{\rm M_{\odot}}$}
\newcommand{\kms}{${\rm{\ }km{\ }s^{-1}}$}
\newcommand{\nbody}{$N$-body}
\newcommand{\Rvir}{$R_{\rm vir}$}
\newcommand{\Mvir}{$M_{\rm vir}$}

\newcommand{\zform}{$z_{\rm form}$}

\newcommand{\Ehost}{$\vec{E}_{1, \rm host}$}
\newcommand{\Rapo}{$\vec{R}_{\rm sat}^{\rm apo}$}
\newcommand{\Lapo}{$\vec{L}_{\rm sat}^{\rm apo}$}

\newcommand{\Table}[1]{Table~\ref{#1}}
\newcommand{\Sec}[1]{Section~\ref{#1}}

\newcommand{\Fig}[1]{Fig.~\ref{#1}}
\newcommand{\mlapm}{\texttt{MLAPM}}

\newcommand{\ea}{et~al.~}                            

\begin{document}


\title{Anisotropy in the Distribution of Satellite Galaxy Orbits}

\author{Alexander Knebe\altaffilmark{1}, 
        Stuart P.D. Gill\altaffilmark{1}, 
        Brad K. Gibson\altaffilmark{1},
        Geraint F. Lewis\altaffilmark{2}, 
        Rodrigo A. Ibata\altaffilmark{3}, and
        Michael A. Dopita\altaffilmark{4}}

\altaffiltext{1}{Centre for Astrophysics~\& Supercomputing, 
                 Swinburne University,Mail\#31, PO Box 218, 
                 Hawthorn, VIC 3122, Australia.}
\altaffiltext{2}{Institute of Astronomy, School of Physics, A29, 
                 University of Sydney, NSW 2006, Australia}
\altaffiltext{3}{Observatoire de Strasbourg, 11, Rue de l'Universite, 
                 F-6700, Strasbourg, France}
\altaffiltext{4}{Research School of Astronomy and Astrophysics, 
                 Australian National University, Weston Creek Post Office, 
                 ACT 2611, Australia}

\begin{abstract}
Nearby clusters such as Virgo and Coma possess galaxy distributions
which tend to be aligned with the principal axis of the cluster
itself. This has also been confirmed by a recent statistical analysis
of some 300 Abell clusters where the effect has been linked to the
dynamical state of the cluster. Moreover, the orbits of satellite
galaxies in galactic systems like our own Milky Way also demonstrate a
high degree of anisotropy -- the so-called Holmberg effect, the origin
of which has been the subject of debate for more than 30 years.  This
study presents the analysis of cosmological simulations focusing on
the orbits of satellite galaxies within dark matter halos.  The
apocentres of the orbits of these satellites are preferentially found
within a cone of opening angle $\sim$40$^\circ$ around the major axis
of the host halo, in accordance with the observed anisotropy found in
galaxy clusters.  We do, however, note that a link to the dynamical
age of the cluster is not well established as both our oldest dark
matter halos do show a clear anisotropy signal.  Further analysis
connects this distribution to the infall pattern of satellites along
the filaments: the orbits are determined rather by the environment of
the host halo than some "dynamical selection" during their life within
the host's virial radius.
\end{abstract}


\section{Introduction} \label{observations}
Observations of the distribution of bright elliptical galaxies within
the Virgo galaxy cluster show a remarkable collinear arrangement
(West~\& Blakeslee 2000).  Moreover, this axis also appears to be part
of a filamentary bridge connecting Virgo and the rich cluster Abell
1367. This phenomenon has already been recognized (Arp 1968;
Bingegeli, Tammann~\& Sandage 1987) but only with accurate
measurements of distance did it become possible to discriminate
between this being a genuine three-dimensional structure or merely a
chance alignment of galaxies. West~\& Blakeslee (2000) based their
distances upon the surface brightness fluctuations method
(cf. Blakeslee, Ajhar~\& Tonry 1999 for a recent review), concluding
that not only were the brightest cluster members distributed
anisotropically over the cluster, but the distribution of dwarf
ellipticals was elongated in the same direction (Bingelli 1999).
Plionis~\ea (2003) investigated this substructure-cluster correlation
statistically using 303 Abell clusters. They affirm that there is
indeed such a signal and also that this signal is related to the
dynamical state (and the environment) of the cluster.

Moreover, even on smaller scales -- in galactic systems -- there are
clear observational indications that the distribution of the orbits of
satellite galaxies is biased towards the galactic pole (Holmberg 1969)
giving an anisotropic distribution.  Zaritsky~\ea (1997) found that
satellites of (isolated) disk galaxies are scattered asymmetrically
about the parent galaxy and aligned preferentially with the disk minor
axis. At first this signal was only prominent for distances out to
$\sim$50~kpc from the host (Holmberg 1969), but an analysis based on a
much larger sample (cf.  Zaritsky~\ea 1993) of satellites extends it
to 200~kpc (Zaritsky~\ea 1997).  This result is confirmed by a study
of the satellites orbiting M31 (Hartwick 2000; Grebel, Kolatt~\&
Brandner 1999).  In addition, for the one galaxy where individual
satellite orbits are known, the Milky Way, there is also evidence that
that the orbits are preferentially polar (Zaritsky~\& Gonzalez 1999).
This is derived from information based upon the alignment of
satellites on the sky (Kunkel~\& Demers 1976; Lynden-Bell 1982), the
orientation of the Magellanic Stream (Mathewson, Clearly~\& Murray
1974), the three-dimensional distribution of satellites (Majewski
1994; Hartwick 1996), and their actual velocities (Scholz~\& Irwin
1994).

All these observations clearly indicate that on scales spanning from
galaxy clusters down to galaxies there is a signal indicating a
correlation between the alignment of substructure and the shape of the
gravitational potential this substructure moves in. However, the
source of this alignment of satellites has been debated for a number
of years, with two potential solutions suggested to explain this
puzzling arrangement. If the distribution of infalling satellites is
initially spherical, then dynamical selection may preferentially
destroy or suppress the star-formation in those on more equatorial
orbits\footnote{With \textit{equatorial} we mean "perpendicular to the
major axis"}. For instance, Pe\~narubbia, Kroupa \& Boily (2002) have
argued that in flattened dark matter halos, the timescale for
dynamical friction is substantially longer for satellites that are on
polar orbits.  Alternatively, the present non-isotropic distribution
of satellite systems may reflect the fact that it has always been
non-isotropic, with the satellites being accreted along preferred
directions.  These hypotheses can be tested in numerical simulations
of structure formation.  However, in previous attempts to decipher
this signal from cosmological \nbody\ simulations there still remains
a certain amount of uncertainty (Zaritsky~\ea 1997). The first fully
self-consistent simulations targeting the subject were performed by
Tormen (1997). They though lacked resolution in time, space and mass
to perform a detailed analysis of the satellite population for
differing environments, and hence more refined simulations are
required. Tormen (1997) could not follow the satellite distribution
within the host's virial radius but rather tracked all progenitors
prior to accretion. This allowed him to analyze the infall pattern
rather than the orbital evolution of the satellites. The aim of this
study is to investigate this subject with the latest state-of-the-art
high-resolution \nbody-simulations. We focus upon a detailed analysis
of the temporal and spatial properties of satellite galaxies residing
within host galaxy clusters that formed fully self-consistently within
a cosmological framework.

\begin{deluxetable}{ccccccccc}
\tablecaption{Summary of the eight host dark matter halos. \label{DMhost}}
\tablewidth{0pt}
\tablehead{ 
\colhead{Halo}                            & 
\colhead{\Rvir [\hkpc]}                   & 
\colhead{$v_{\rm circ}^{\rm max}$[\kms]}  & 
\colhead{\Mvir [$10^{14}$\hMsun]}         & 
\colhead{\zform}                          & 
\colhead{age [Gyr]}                       & 
\colhead{$N_{\rm sat}(<\!r_{\rm vir})$}   & 
\colhead{$T$} 
}
\startdata
 \# 1 &  1349 & 1125 & 2.87 & 1.16 & 8.30 & 158 & 0.67 \\
 \# 2 &  1069 &  894 & 1.42 & 0.96 & 7.55 &  63 & 0.87 \\
 \# 3 &  1081 &  875 & 1.48 & 0.87 & 7.16 &  87 & 0.83 \\
 \# 4 &   980 &  805 & 1.10 & 0.85 & 7.07 &  57 & 0.77 \\
 \# 5 &  1356 & 1119 & 2.91 & 0.65 & 6.01 & 175 & 0.65 \\
 \# 6 &  1055 &  833 & 1.37 & 0.65 & 6.01 &  85 & 0.92 \\
 \# 7 &  1014 &  800 & 1.21 & 0.43 & 4.52 &  59 & 0.89 \\
 \# 8 &  1384 & 1041 & 3.08 & 0.30 & 3.42 & 251 & 0.90 \\
\enddata

\end{deluxetable}

\section{The Simulations} \label{simulations}
There is mounting  evidence that, despite its problems,  the Cold Dark
Matter  structure  formation   scenario  provides  the  most  accurate
description of  our Universe. Observations point  towards the standard
\LCDM\ according to which the universe is comprised of about 28\% dark
matter, 68\% dark energy, and luminous baryonic matter (i.e. galaxies,
stars, gas, and dust) at a mere 4\% level (cf. Spergel~\ea 2003). This
so-called "concordance model" induces hierarchical structure formation
in  which small  objects form  first  and subsequently  merge to  form
larger  objects.

The analysis presented in this study is based upon a suite of eight
high-resolution \nbody\ simulations within the \LCDM\ concordance
cosmology.  They were carried out using the publicly available
adaptive mesh refinement code \mlapm\ (Knebe, Green~\& Binney 2001).
\mlapm\ reaches high force resolution by refining all high-density
regions with an automated refinement algorithm.  The refinements are
recursive: the refined regions can also be refined, each subsequent
refinement having cells that are half the size of the cells in the
previous level.  This creates an hierarchy of refinement meshes of
different resolutions covering regions of interest.  The refinement is
done cell-by-cell (individual cells can be refined or de-refined) and
meshes are not constrained to have a rectangular (or any other)
shape. The criterion for (de-)refining a cell is simply the number of
particles within that cell and a detailed study of the appropriate
choice for this number as well as more details about the particulars
of the code can be found elsewhere (Knebe, Green~\& Binney, 2001).

Each run focuses on the formation and evolution of one particular dark
matter halo containing of order greater one million particles. The
force resolution reached by \mlapm\ in the high-density regions is
2\hkpc\ corresponding to roughly 0.25\% of the host's virial radii
\Rvir. Following Lacey~\& Cole (1994) we define the formation redshift
\zform\ as the time when the halo contains half of its mass today.
The host halos were carefully chosen to sample a variety of different
triaxialities and merger histories as summarized in
\Table{DMhost}. The virial radii given in that Table correspond to the
point where the density of the host (measured in terms of the
cosmological background density $\rho_b$) drops below $\Delta_{\rm
vir}=340$ with \Mvir\ being the mass enclosed by that sphere. All
satellite galaxies orbiting in and around the host halo are identified
using the \mlapm-based halo finder described in a companion paper
(Gill, Knebe~\& Gibson 2004). This new halo finder is based upon the
\mlapm\ grid structure and works at the same resolution level as the
\nbody\ code itself thus assuring not to miss or include any objects
not resolved in the actual simulation. The satellites are then
individually traced from \zform\ onwards until redshift $z=0$.  The
mass spectrum of those satellites accounted for in the following
analysis can be described by a declining power-law $dn/dM\propto
M^{-\alpha}$ with $\alpha \approx 1.7-1.9$ in the range from
$2\times10^{10}$\hMsun\ (applied mass-cut corresponding to 100
particles and explaining the rather 'low' number for $N_{\rm
sat}(<\!r_{\rm vir})$ in \Table{DMhost}) up to
$\sim10^{13}$\hMsun. The total fraction of mass locked up in
satellites never exceeds $\leq$15\% though.

The high temporal sampling of the outputs (i.e. $\Delta t \approx 0.2$
Gyrs) allows us to accurately measure the orbital parameters of the
satellite galaxies and a detailed analysis of their disruption and
survival history will be presented elsewhere (Gill, Knebe~\& Gibson
2004). Gill~\ea show that the distribution of orbital eccentricities
for the satellite population peaks about the value $\epsilon \approx
0.6$ where $\epsilon = 1-p/a$ is defined using the last pericenter $p$
and apocenter $a$ of its orbit. The pericenter distribution itself
peaks about 35\% of \Rvir\ for all host halos and most of the
satellites (i.e. more than 75\%) had had one full orbit with the
maximum number of orbits found to be 3--5 depending on the host
halo. The angular momentum vector of the satellite \Lapo=\Rapo$\times
\vec{V}_{\rm sat}^{\rm apo}$ was calculated at its last apocenter
passage \Rapo.

In order to probe the alignment with the shape and orientation of the
host halo we calculated the eigenvectors $\vec{E}_{\rm 1,2,3}$ (with
$\vec{E}_{\rm 1}$ being the major axis) of its inertia tensor
using only the "core" region as defined by the $6^{\rm th}$
refinement level in \mlapm, i.e. the boundary of this refinement level
is an isodensity contour. According to the refinement criterion
adopted in the simulations the $6^{\rm th}$ level surrounds material
about 3000 times denser than $\rho_b$ or in other words 9 times denser
than the material at the virial radius.  The host halos are well
described by the density profiles advocated by Navarro, Frenk~\& White
(1997) with concentration in the range $c = 5-7$. Therefore,
a density of roughly $9\times \rho(r_{\rm vir})$ corresponds to about
the half-mass radius of the host.  The eigenvectors now define the
orientation of the host halo and the coordinate system used to measure
the orbits of the satellites, respectively. Moreover, its eigenvalues
$a>b>c$ can be used to construct the triaxiality parameter
$T=(a^2-b^2)/(a^2-c^2)$ (Franx, Illingworth~\& Zeeuw 1991) presented
in \Table{DMhost}, too.

The left panel of \Fig{NdotE} presents the (normalized) distribution
of angles between \Rapo\ and \Ehost.  This graph shows that there is a
clear trend in at least six of the eight halos for the two vectors to
be aligned (the distribution peaks at $0^\circ$ and $180^\circ$,
respectively) meaning that the orbits of the satellites are
preferentially found along the major axis of the host.  The right
panel of \Fig{NdotE} shows the (normalized) distribution of angles
between the angular momentum vector \Lapo\ of the satellite at its
last apocenter and \Ehost, clearly revealing evidence for these two
vectors to be parallel (distribution peaks at $90^\circ$).  We like to
stress that only satellite galaxies that at least had one or more
complete orbits were taken into account in \Fig{NdotE}; the figure is
\textit{not} based upon the infall pattern of satellites as
investigated by, for instance, Tormen (1997). We also need to stress
that the distributions presented in that Figure are normalized,
i.e. they are corrected for the bias introduced by plotting them as a
function of the angle $\theta$ rather than $\cos\theta$. The data is
binned equally spaced in angle ranging from 0$^\circ$ to 180$^\circ$,
and therefore the area probed on the sphere varies with
$\theta$. Therefore the distribution needed to be normalized by the
the respective area $A$ specified by the actual range of angles
[$\theta-\Delta\theta/2,\theta+\Delta\theta/2$]. This area is
proportional to $A \propto \cos(\theta-\Delta\theta/2)-\cos(\theta+\Delta\theta/2)$.

\section{Conclusions and Discussion} \label{conclusions}
\begin{figure}
\plottwo{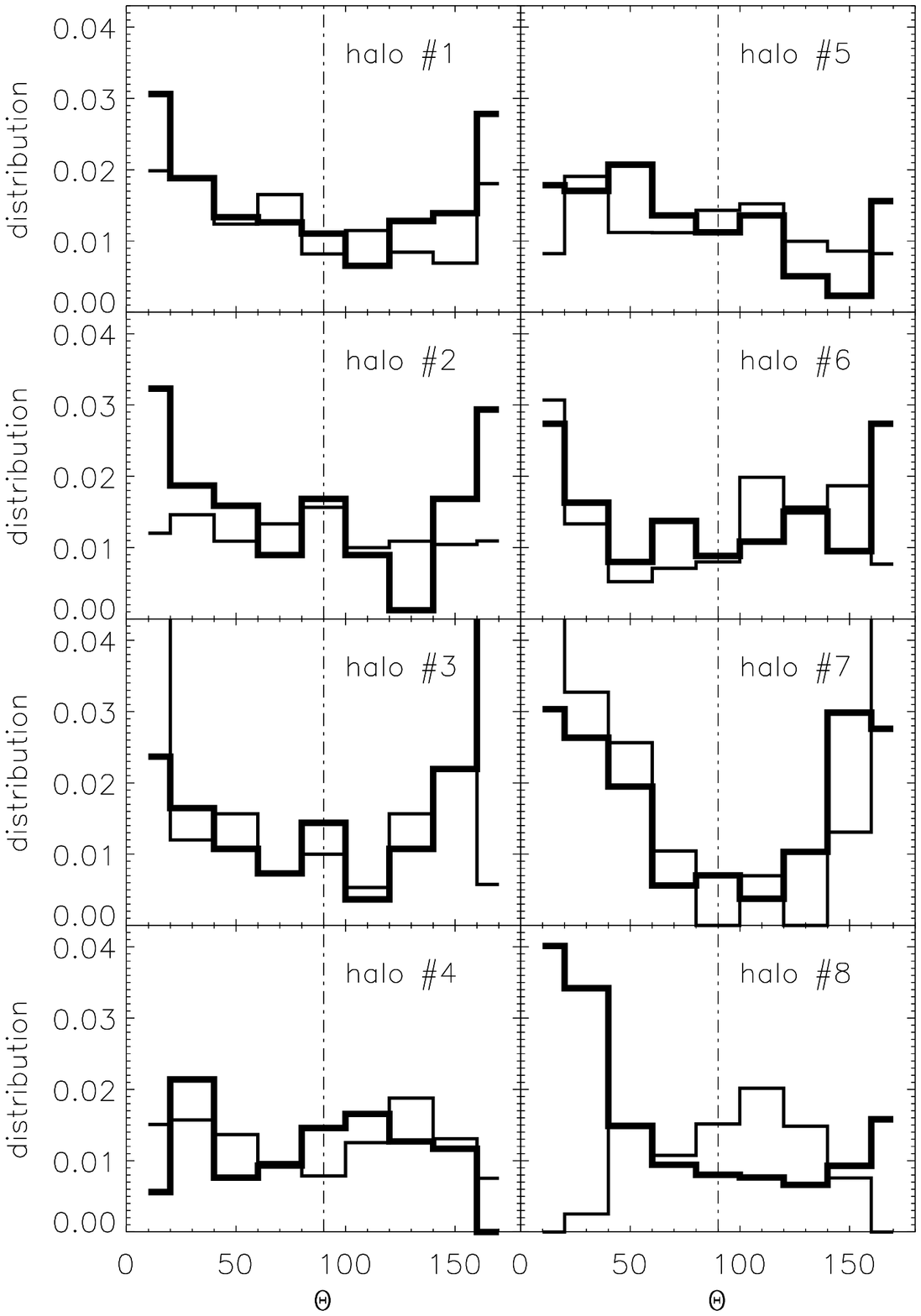}{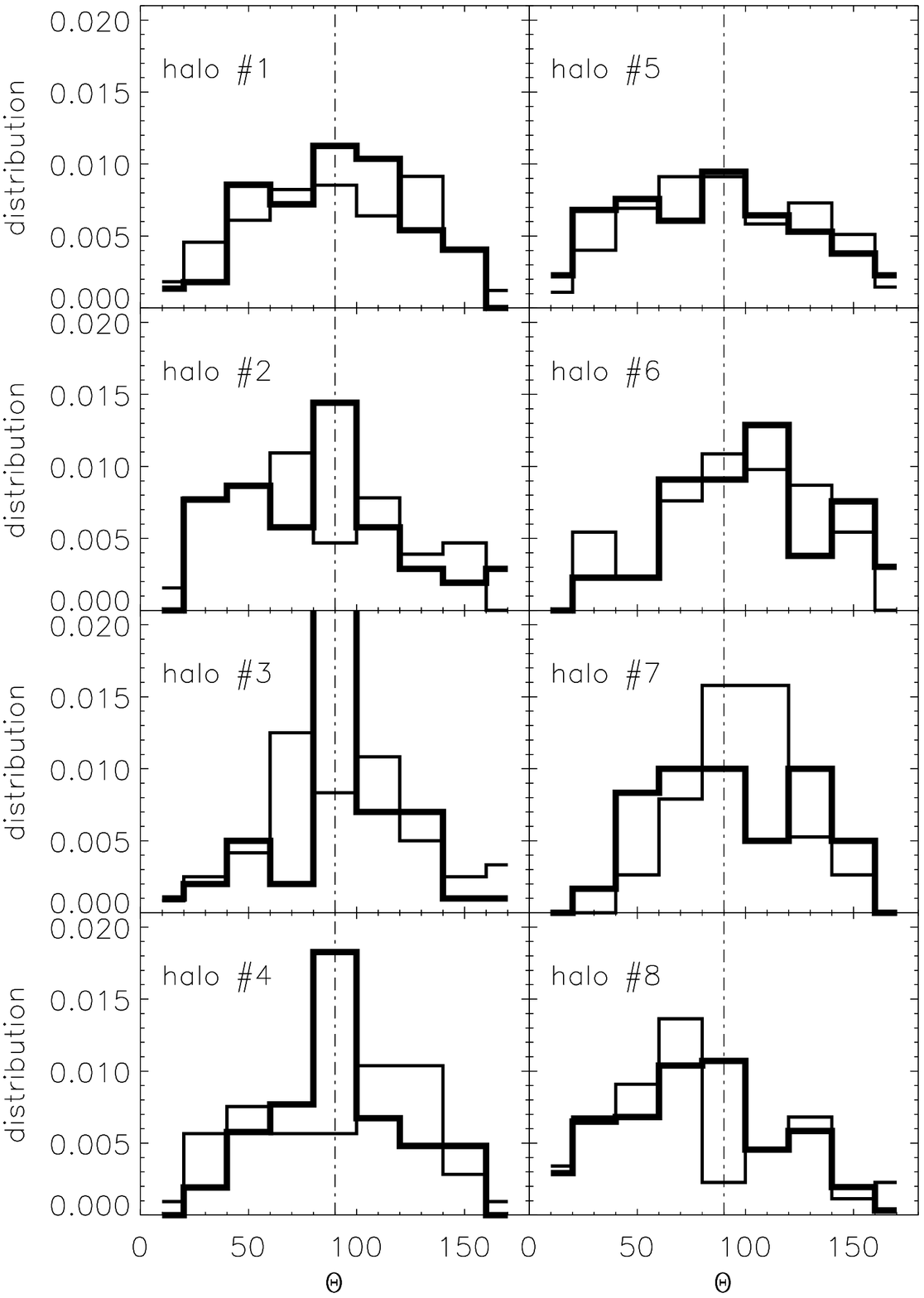}
\caption{The  left  panel shows  the  (normalized) distribution  of angles  between
         position vector \Rapo\ of the satellites measured at the last
         apocentre and  the major axis \Ehost\ of  the halo.  $\theta$
         values of  0$^\circ$ and 180$^\circ$  indicate alignment with
         the the  principal axis of  the host.  The right  panel shows
         the  (normalized) distribution of angles  between angular  momentum vector
         \Lapo\ of  satellites measured at the last  apocentre and the
         major  axis  \Ehost\  of  the  halo.   A  $\theta$  value  of
         90$^\circ$  means that  the  orbital plane  of the  satellite
         contains  the  major  axis  of  the host  where  $\theta$  of
         0$^\circ$ and  180$^\circ$ would require for the  plane to be
         perpendicular  to  the   host's  principal  axis.  The  thin
         histograms are for disrupted satellites. In both cases only
         satellites more massive than $2\times 10^{10}$\hMsun\ 
         (100 particles) that had one full orbit were taken into account.
         \label{NdotE}}
\end{figure}

There are two possible scenarios that explain the substructure-cluster
alignment found in observations of galaxy clusters (Plionis~\ea 2003;
West~\& Blakeslee 2000) and the analysis of \nbody-simulations
presented in this study.  Either some dynamical process can be held
responsible, meaning that initially the orbits were randomly
distributed and only some satellites survive to give rise to the
observed correlation in \Fig{NdotE}, or alternatively, their orbital
parameters are imprinted upon them at the time they enter the host
halo as already pointed out by Tormen (1997).

``Dynamical destruction'', whereby satellites on more
equatorial orbits suffer more tidal disruption, can be rejected because a
comparable (although not as pronounced) alignment signal is found when
restricting the analysis to those satellites that are in fact
disrupted: We placed mass-less tracer particles at the last centre of
the satellites before they were classified "disrupted".  These
``disrupted systems'' are presented as the thin lines in \Fig{NdotE}.

\begin{figure}
\plotone{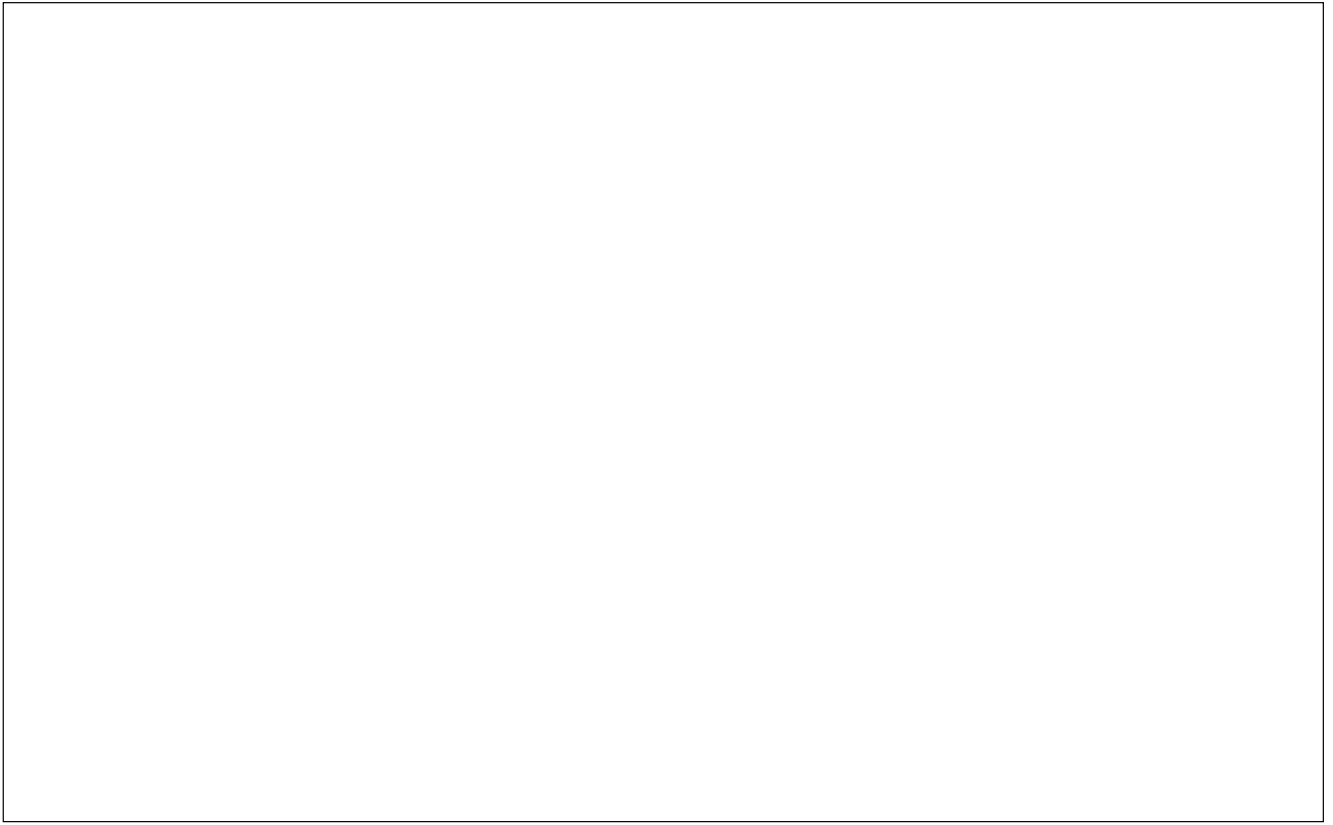}
\caption{Orbits of all satellite galaxies identified at formation time (dark) 
         to redshift $z=0$ (light). The left panel shows how the host halo
         fits into the surrounding large-scale structure presented at formation 
         time. The right panel zooms into the region marked in the left panel, 
         this time not showing low-resolution particles. The shape (and position) 
         of the underlying host halo at redshift $z=0$ is indicated by the 
         best-fit ellipse to the grid used to calculate its triaxiality.
         The vertical line extending to top edge of panel indicates the
         principal axis of the host. With the assumptions presented in this 
         paper, the pole of the host galaxy within the dark matter halo is 
         aligned with this principle axis.
         \label{SatelliteOrbits}}
\end{figure}

We therefore conclude that it must be the initial distribution of the
satellite systems that is responsible for the present day alignment.
For example, galaxies are "funneling along the filaments"
(Kitzbichler~\& Saurer 2003) which has also been confirmed by X-ray
observations of the spatial distribution of substructure in galaxy
clusters (West, Jones~\& Forman 1995) and biases the types of possible
orbits.  It would now be reassuring to confirm a link between the host
halo shape and the surrounding environment, i.e. are the filaments
that feed the halo with material preferentially angled with respect to
the orientation of the host halo?  When projecting the positions of
the satellites onto a sphere at the time they enter the virial radius
of the host it is clear that they are not randomly distributed; they
cluster in directions linked to the filamentary structure surrounding
the host halo, as found before observationally (e.g.  West~\&
Blakeslee 2003; Plionis~\& Basilikos 2002, and references therein) and
in \nbody-simulations (e.g.  Faltenbacher 2002; Hatton~\& Ninin 2001;
Onuora~\& Thomas 2000; Colberg~\ea 1999; Splinter~\ea 1997; Tormen
1997).  Tormen (1997) already pointed out that there is a strong
alignment between the distribution of infalling satellite galaxies and
the shape of the dark matter host. However, Tormen's simulations did
not have the spatial and mass resolution to investigate the orbits of
satellites \textit{within} the host halo. We are, for the first time,
tracing extremely well-resolved dynamics of the satellites within the
dark matter hosts confirming that the alignment for both, disrupted
and surviving, satellites is maintained for several orbital
periods. \Fig{SatelliteOrbits} now presents a visual impression of the
trajectories of all surviving satellites identified at formation time
of host halo~\#3 until redshift $z=0$, qualitatively supporting the
scenario that these orbits are linked to the filamentary structure
which will leave its imprint in observed anisotropy.

Plionis~\ea (2003) also established a link between the alignment and
the dynamical age of the clusters. Our simulations though show that
this finding can not be generalized: both our oldest halos, in
which the satellites had as many as 4--5 orbits, do show the
correlation signal.  We do not observe a trend for a "randomization"
of the orbits in older halos. The satellites actually preserve the
alignment with the host they had when they first fell into the
cluster.

In \Sec{observations} we showed that the (observational)
substructure-alignment signal spans from galaxy clusters down to
galactic scales.  Our results are mostly applicable to the former, and
any extrapolation to smaller scales has to be handled with care. In
the case for galaxies, for instance, the interaction between the
galactic disk and the incoming satellite could enhance destruction in
the galaxy plane, i.e. satellite on prograde orbits decay faster than
the ones on retrograde (or polar) orbits due to orbital resonances
between the disk and the satellites (Pe\~narubbia, Kroupa \& Boily
2002).  Based on theoretical and numerical studies (cf. Lacey~\& Cole
1993 and Moore~\ea 1999) it should be possible to re-scale our
simulations to a Milky Way sized object by requiring that the maximum
of the circular velocity curves of our halos equals 220~\kms.  The
scaling factor lies in the range of about 4--5 (cf.~$v^{\rm max}_{\rm
circ}$ values in \Table{DMhost}) and hence our ``rescaled dark matter
halos'' would correspond to the ones in the observational data with
virial radii in the range 260--295\hkpc. This scaling factor in length
entails a scaling in mass of 64--125 (simply the length scale to the
power of three). This brings the mass of our galaxy clusters down to
$\sim 10^{12}$\hMsun\ which agrees with the dark matter mass inferred
for our Milky Way (Freeman 1996). More problematic, however, are the
ages of our systems: our halos are only $\leq$8.3~Gyrs old opposed to
12~Gyrs for the Milky Way. Satellites in the Milky Way had the chance
to complete nearly twice as many orbits leaving more space for an
explanation based upon the dynamical destruction scenario.

When re-scaling our data, putting aside the age issue, and trying to
explain the Holmberg effect, another uncertainty comes into play: the
orientation of the stellar disk. In a triaxial potential there are in
general stable closed orbits about both the major and the minor axis
(e.g., Binney \& Tremaine 1987). So, in principal the disk plane could
either be perpendicular to the major or perpendicular to the minor
axis. Under the assumption it lies perpendicular to the major axis of
the dark matter halo the satellites in our simulations will be on
polar orbits. There are indications that this configuration results in
the most stable disk configuration within a triaxial halo (Hayashi~\ea
2003). Moreover, even though there are clear indications that the
angular momentum of the dark matter is well aligned with the minor
axis of the halo (e.g. Warren~\ea 1992), van den Bosch~\ea (2002)
showed that the angular momentum of the baryonic component (i.e. gas)
not necessarily follows that of the dark matter distribution. They
found an average misalignment between $\vec{L}_{\rm gas}$ and
$\vec{L}_{\rm DM}$ of the order of 40$^\circ$ in their numerical
simulations. Hence the orientation of the galactic disk with respects
to the dark matter halo is not well determined. Turning to
observations does not resolve this question, with studies of polar
rings indicating strongly oblate halos (i.e.  Iodice~\ea 2003), while
others suggest dark matter halos are more spherical, or even oblate
(Olling \& Merrifield 2000; Ibata~\ea 2001). The situation is not at
all clear, but the assumption that disks are perpendicular to the
major axis (which is in agreement with results presented by
Hayashi~\ea 2003) would provide an explanation for the Holmberg
effect, even though this is a very speculative interpretation.

\acknowledgments
The simulations presented in this paper were carried out on the
Beowulf cluster at the Centre for Astrophysics~\& Supercomputing,
Swinburne University.  The support of the Australian Research Council
and the Swinburne Research Development Grants Scheme is gratefully
acknowledged.  GFL thanks Suede for ``Introducing the band''.

\end{document}